\documentclass[pra,twocolumn,10pt,superscriptaddress,showpacs]{revtex4-1}
\usepackage{amsmath,graphicx,amsfonts,bm,amssymb}
\usepackage{times}

\begin{document}

\title{Robust interface between flying and topological qubits}

\author{Zheng-Yuan Xue}  \email{zyxue@scnu.edu.cn}

\affiliation{Department of Physics and Center of Theoretical and Computational Physics, The University of Hong Kong, Pokfulam Road, Hong Kong, China}

\affiliation{Guangdong Provincial Key Laboratory of Quantum Engineering and Quantum Materials, and School of Physics\\ and Telecommunication Engineering, South China Normal University, Guangzhou 510006, China}

\author{Ming Gong}  \email{skylark.gong@gmail.com}
\affiliation{Department of Physics and Center for Quantum Coherence, The Chinese University of Hong Kong, Shatin, N.T., Hong Kong, China}

\author{Jia Liu}
\affiliation{Department of Physics and Center for Quantum Coherence, The Chinese University of Hong Kong, Shatin, N.T., Hong Kong, China}

\author{Yong Hu}
\affiliation{Department of Physics and Center of Theoretical and Computational Physics,  The University of Hong Kong, Pokfulam Road, Hong Kong, China}
\affiliation{School of Physics, Huazhong University of Science and Technology, Wuhan 430074, China}

\author{Shi-Liang Zhu}
\affiliation{National Laboratory of Solid State Microstructure and School of  Physics, Nanjing University, Nanjing 210093, China}
\affiliation{Synergetic Innovation Center of Quantum Information and Quantum Physics, University of Science and Technology of China, Hefei 230026, China}

\author{Z. D. Wang} \email{zwang@hku.hk}
\affiliation{Department of Physics and Center of Theoretical and Computational Physics, The University of Hong Kong, Pokfulam Road, Hong Kong, China}

\date{\today}

\begin{abstract}
\textbf{Hybrid architectures, consisting of conventional and topological qubits, have recently attracted much attention due to their capability in consolidating  robustness of topological qubits and universality of conventional qubits. However, these two kinds of qubits are normally constructed in significantly different energy scales, and thus the energy mismatch is a major obstacle for their coupling, which can support the exchange of quantum information between them. Here we propose a microwave photonic quantum bus for a strong direct coupling between the topological and conventional qubits, where the energy mismatch is compensated by an external driving field. In the framework of tight-binding simulation and perturbation approach, we show that the energy splitting of Majorana fermions  in a finite length nanowire, which we use to define topological qubits, is still robust against local perturbations due to the topology of the system. Therefore, the present scheme realizes a rather robust interface between the flying and topological qubits. Finally, we demonstrate that this quantum bus can also be used to generate multipartitie entangled states with the topological qubits.}
\end{abstract}

\maketitle


Recently, topological quantum computation \cite{Moore,rg,kitaev01,tqc,tqc2} has been resurfaced due to the invention of an experimental accessible  way on the realization of Majorana fermion (MF) --- a self-conjugate fermion who obeys non-Abelian exchange statistics \cite{Ivanov}. For the past years, this kind of exotic particles have been predicted to exist in the $\nu = 5/2$ fractional quantum Hall state \cite{Moore}, vortex core in two dimensional chiral \emph{p}-wave superconductor \cite{rg}, and one dimensional (1D) nanowire in proximity to a \emph{p}-wave superconductor \cite{kitaev01}. However, none of them have readily be used for the realization of MFs. Remarkably, it was indicated that the unconventional \emph{p}-wave pairing can be induced by coupling  the spin-orbit  interaction to a conventional $s$-wave pairing \cite{FuLiang, sau, JLiu}. Along this line, several theoretical schemes based on one-dimensional systems have been proposed \cite{1d1,1d2,gate,Zhao1,Zhao2}, and  experimental investigations  of possible MFs have also been made \cite{Mourik12, Rokhinson12,Das12,Deng12}, making the MFs be a kind of promising candidate for implementing topological quantum computation \cite{alicea12,st,stern, Beenakker13, Flensberg12}.

Unfortunately, braiding operations of MFs are not universal for quantum computation because only a few quantum gates can be obtained. One possible alternative scenario is to use the hybrid architecture between topological and conventional qubits, which can consolidate the advantages of both systems --- the topological qubits are robust against perturbations while the conventional qubits can be used to perform universal quantum computation via coherent control. So far, many schemes have been proposed to interface topological and conventional qubits \cite{Hassler10,Hassler11,CYHou11,d1,jiang,Bonderson11,Leijnse121,Leijnse122,zhangzt,s4}, with most being used to measure the topological qubits.  Generally, there is essentially an obstacle for the realization of strong coupling between conventional and topological qubits, that is, the energy mismatch effect. The topological qubits are constructed in a degenerate zero energy subspace, while conventional qubits are usually defined by two isolated energy levels with different energy, which is essential for coherent operations via Rabi oscillation. Therefore, direct interfacing that admits the energy exchange between different qubits is not allowed. Meanwhile, in order to couple long distance qubits, a photonic quantum bus for topological qubits is of significant importance, where errors from these hybrid architectures can be corrected for a much higher threshold ($\varepsilon \sim 0.14$) \cite{imp1,imp2}, which has already been achieved \cite{Ladd10}.  However, for topological qubits couple to a cavity mode, only the induced energy shift, due to the large energy mismatch effect,  has been investigated before \cite{Schmidt131,Schmidt132,ZYXue13,Hyart13,Cottet13,Muller13,Ginossar13}.

Here we propose a microwave photonic quantum bus for strong coupling between conventional and topological qubits  in a circuit QED scenario \cite{Schoelkopf131,Schoelkopf132}. We use MFs in a finite length nanowire  \cite{Kovalev14} and an ac driving field  to compensate the energy mismatch between the MFs and cavity frequency \cite{pac,jun,jl,kt}. It is noted that a similar setup based on dc driven has been employed in Ref. \cite{ohm}, where the same interaction Hamiltonian is obtained based on dipole approximation of the topological qubit and treat the semi-classical dynamics of the coupled system. In realistic experiments, the dc bias may displace the working point of the qubit off its optimal point, which enhances the charge sensitivity of the quantum device. Therefore, to investigate the quantum dynamics of similar systems,  dc driven will introduce additional charge noise \cite{sci886,sq}. This problem can be avoided using the ac bias, in which the averaged bias is zero in a full period. We therefore expect that the ac bias can lead to a better performance in our model focusing on quantum dynamics. Then, using the tight-binding simulation and second-order perturbation theory, we show that the energy splitting of MFs in a finite length nanowire, which we use to define topological qubits, is robust against local perturbations. This robustness is ensured by the topology of the system, in which, although the perturbations may induce the coupling between edge states and other extended wave functions, their contribution to the splitting energy are almost cancelled. Thus our scheme realizes a robust interface between flying and topological qubits.  Finally, we show that this quantum bus can be used to generate multipartite entangled states with the topological qubits, which are impossible by braiding of MFs.

\begin{figure}[tb]
\begin{center}
\includegraphics[width=3.0in]{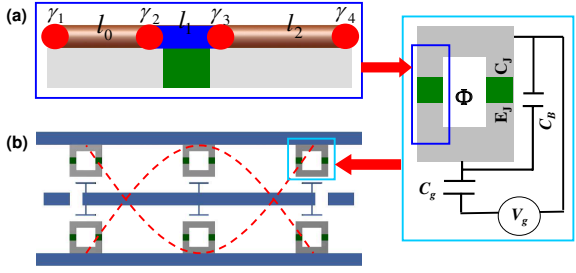}
\end{center}
\caption{\textbf{The proposed setup.} (a) Topological qubits encoded by 4 MFs (red filled dots) locate at the intersections of the topological trivial (blue) and nontrivial phases of a semiconducting nanowire, which is deposit on top of a transmon qubit (right panel). The transmon qubit is controlled by a dc gate voltage $V_\text{g}$ via a gate capacitor $C_\text{g}$, from which an ac voltage bias can also be introduced. The Josephson junctions  of the qubit have capacitance $C_\text{J}$ and Josephson coupling energy $E_\text{J}$,  which are shunted by a large capacitance $C_\text{B}$. (b) The full-wave section of a 1D transmission line resonator (cavity), where the transmon qubits are located at the antinodes of the cavity mode and interact to it by the capacitive coupling.}
\label{qubit}
\end{figure}

\bigskip

\noindent\textbf{Results}\\
\textbf{Interfacing topological and flying qubits.}
We first introduce our setup to realize strong coupling between topological and
flying qubits, as schematically shown in Fig. \ref{qubit}.  We consider a
spin-orbit coupled semiconductor (InAs or InSb) nanowire deposited on a
superconducting transmon qubit \cite{sq}. Topologically protected MFs, defined as $\gamma_1$ to $\gamma_4$ from left to right, can be realized when the nanowire is driven to the topological phase regime \cite{1d1}.  In particular, due to the presence of the MFs  $\gamma_2$ and $\gamma_3$, the single electron  tunnelling across the junction will also appear besides the usual Cooper pair tunnelling. Moreover, since the difference between resonant energies of the two type of tunnelings is sufficiently large, it is reasonable to assume that only one of them can be resonantly addressed by the biased voltage.

For a finite nanowire, the coupling between the MFs leads to an energy splitting. In this case, the MFs gain a finite energy while their wave functions are still well localized at the two ends. Roughly speaking,  we still have $\gamma \simeq \gamma^\dagger$, thus these states with nonzero energies can still be used to encode information for topological quantum computation. For the four MFs defined in Fig. \ref{qubit}(a), we assume their distances to be $l_0$,  $l_1$ and $l_2$, respectively, which are much longer than the Cooper pair coherent length. In this case, the Hamiltonian of the MF system reads
\begin{eqnarray} \label{mf}
H_{\text{MF}}  =  i {E_{1} \over 2}
\gamma_1\gamma_2 + i {E_{2} \over 2} \gamma_3\gamma_4 + i {E_\text{M}
\over 2} \cos\left({\phi \over 2}\right) \gamma_2\gamma_3,
\end{eqnarray}
where $\phi$ is the phase difference across the junction, $E_1$,
$E_2$  and $E_\text{M}=4\sqrt{D}\Delta$ are the coupling strength with
$D$ being the transmission probability of the junction. Usually, to maintain stable topological protection, the MF splitting energy $E_\text{i}$ ($\sim$ MHz to 0.1 GHz) is much smaller than the microwave cavity frequency (3 - 10 GHz). This large energy mismatch prohibits the direct resonant coupling between these two distinct systems. To overcome this shortcoming, we propose to use a microwave bias voltage to match the energy difference.  In this way, the phase difference $\phi$ consists of three contributions: the difference between the two superconductors $\varphi$,  the microwave driven field $V_\text{RF}=A \sin \omega t$  and the capacitively coupling to the quantized cavity field, that is, $V_\text{c}=V_0\left(a e^{-i\omega_\text{c} t}+a^\dag e^{i\omega_\text{c} t}\right)$. As the total induced bias voltage for the junction is $V_\text{b}=\beta
(V_\text{RF}+ V_\text{c})$ with $\beta=C_\text{g}/C_{\Sigma}$ and $C_\Sigma=C_\text{g}+C_\text{B}+2C_\text{J}$, the phase difference is given by
\begin{eqnarray}
\phi/2&=& \varphi/2 + e\int_0^t V_\text{b}  dt  \\
&=&\frac{\lambda_\text{c}}{\omega_\text{c}}
\left(a e^{-i\omega_\text{c} t}+a^\dag e^{i\omega_\text{c} t}\right)
-{\lambda_\text{g} \over \omega} \cos\omega t+  \varphi/2 -\varphi_0, \notag
\end{eqnarray}
where $\lambda_\text{g}=e\beta A$, $\lambda_\text{c}=e\beta V_0$ and
$\varphi_0$ is a constant of integration related to the initial
phase difference of the two superconductors. We treat the transmon qubit classically and absorb $\varphi$ into $\varphi_0$. Normally, $\lambda_\text{c}/\omega_\text{c} \ll 1$, and thus  we can handle the MF Hamiltonian perturbatively. Up to the leading order we obtain
\begin{eqnarray} \label{total3}
H _{\text{MF}} &=& i {E_{1} \over 2} \gamma_1\gamma_2
+ i {E_{2} \over 2} \gamma_3\gamma_4 + i {E_\text{M} \over 2} \cos\left(\theta \cos\omega t + \varphi_0 \right) \gamma_2\gamma_3 \nonumber\\
&+& ig_0 \sin\left(\theta \cos\omega t+ \varphi_0 \right)
\left(a e^{-i\omega_\text{c} t}+a^\dag e^{i\omega_\text{c} t}\right) \gamma_2\gamma_3,
\end{eqnarray}
where $\theta={\lambda_\text{g} /\omega}$ and  $g_0=E_\text{M}\lambda_\text{c}/(2\omega_\text{c})$.

To proceed,  we now construct the conventional Dirac fermion via two MFs,
$c_\text{i,j} =\left(\gamma_\text{i}+i\gamma_\text{j}\right) /\sqrt{2}$. The
eigenstates of $\hat{n}_\text{i,j}=c_\text{i,j}^{\dag}c _\text{i,j}$ define a two
fold degenerate Hilbert space, where $\hat{n}_\text{i,j} = 0, 1$ labels parity of the ground state parity. In the odd parity space, a topological qubit is encoded as
$|0\rangle_\text{t}=|0\rangle_{1,2}|1\rangle_{3,4} $ and
$|1\rangle_\text{t}=|1\rangle_{1,2}|0\rangle_{3,4}$, while the similar
encoding in the even parity subspace is discussed in Ref. \cite{Zhuprl2011}. In this odd parity subspace, we have
\begin{equation} \label{define}
i\gamma_2\gamma_3\rightarrow -\sigma^\text{x}, \quad
i\gamma_1\gamma_2\rightarrow -\sigma^\text{z}, \quad
i\gamma_3\gamma_4\rightarrow \sigma^\text{z}.
\end{equation}
Thus we can express the Hamiltonian in Eq. (\ref{total3}), using Pauli matrices $\sigma^{\text{x, y, z}}$, as
\begin{eqnarray}\label{total4}
H_{\text{MF}} &=&{E \over 2} \sigma^\text{z} -{E_\text{M} \over 2}
\cos\left(\theta \cos\omega t+ \varphi_0 \right)\sigma^\text{x}\notag\\
&&-g_0 \sin \left(\theta \cos\omega t + \varphi_0 \right)
\left(a e^{-i\omega_\text{c} t}+a^\dag e^{i\omega_\text{c} t}\right) \sigma^\text{x},
\end{eqnarray}
where $E=E_2-E_1$. We first consider the time-dependent driven term of the above Hamiltonian, i.e., the $E_\text{M}$ term. The net effect of this term can be modeled as a modulation of the coefficient $E_\text{M}$ when $J_\text{n}(\theta)E_\text{M}/(n\omega)\ll 1$. When $\varphi_0=\pi$, this condition can be fulfilled by choosing $\omega/E_\text{M}=10$ (see the Method section for details). For the single-photon assisted resonate coupling, the Hamiltonian in Eq. (\ref{total4})  reduces to
\begin{eqnarray}\label{total6}
H_{\text{MF}}&=&{E \over 2} \sigma^\text{z}
+{E_\text{M} \over 2} J_0\left(\theta\right)\sigma^\text{x}\notag\\
&&+2g_0 J_1\left(\theta\right) \cos\omega t \left(a e^{-i\omega_\text{c}
t}+a^\dag e^{i\omega_\text{c} t}\right)\sigma^\text{x}.
\end{eqnarray}
In the eigenbasis of the topological qubit, the above Hamiltonian reduces to
\begin{eqnarray}\label{total7}
H_{\text{MF}}&=&{\omega_\text{tq} \over 2} \sigma^\text{z}
+2g_0 J_1\left(\theta\right) \cos\omega t
\left(a e^{-i\omega_\text{c} t}+a^\dag e^{i\omega_\text{c} t}\right)\notag\\
&& \times \left(\sin\vartheta\sigma^\text{z}+\cos\vartheta\sigma^\text{x}\right),
\end{eqnarray}
where $\omega_\text{tq}=\sqrt{E^2+[J_0(\theta)E_\text{M}]^2}$, $\cos\vartheta=E/\omega_\text{tq}$,  and $\sin\vartheta=J_0(\theta)E_\text{M}/\omega_\text{tq}$. Obviously, since $\omega_\text{c}-\omega_\text{tq}\gg g_0$, any direct energy exchange coupling between the two type of qubits is impossible in the absence of the bias. This is expected from our analysis in the introduction.

However, when the energy mismatch between the cavity  field and
the topological qubit is compensated by the bias field, i.e.,
$\omega_\text{c}=\omega+\omega_\text{tq}$, a parametric resonant coupling can be
induced. This also means that the coupling between MFs and
resonator can be switched on/off very easily by controlling the
frequency of the bias field. To see this, we transform
the interaction Hamiltonian in Eq. (\ref{total7}) into the
interaction picture with respective to the qubit Hamiltonian
$H_\text{tq}= \omega_\text{tq}  \sigma^\text{z}/2$, the effective interaction
Hamiltonian reads
\begin{eqnarray}
H_{\text{eff}} &=&\exp(iH_\text{tq}t)(H_{\text{MF}}-H_\text{tq})\exp(-iH_\text{tq}t)\notag\\
&=& g \left(a\sigma^+ + a \sigma^- e^{-2i\omega_\text{tq} t} +h.c.\right)\notag\\
&+&  g \left[a\sigma^+ e^{-i(\omega_\text{c}+\omega-\omega_\text{tq}) t}
+a \sigma^- e^{-i(\omega_\text{c}+\omega+\omega_\text{tq}) t}+h.c.\right]\notag\\
&+& g' \left[a \sigma^\text{z} \left(e^{-i\omega_\text{tq} t}
+e^{-i(\omega+\omega_\text{c}) t}  \right)+ h.c. \right],
\end{eqnarray}
where  $g=g_0 \cos\vartheta J_1\left(\theta\right)$ and $g'=g_0 J_1\left(\theta\right) \sin\vartheta$.   To obtain the maximum coupling strength, one should set $\cos\vartheta=1$, which can be fulfilled when $J_0(\theta)=0$ ($\theta \simeq 2.4$).  In this case, $\omega_\text{tq}=E$, $g\approx g_0/2$ and $g'=0$. Neglecting the oscillating terms using the rotating wave approximation, which is valid when $\omega_\text{tq} \gg g$, the effective Hamiltonian reduces to
\begin{eqnarray} \label{jc}
H_{\text{JC}} = g \left(a\sigma^+ + a^\dagger \sigma^-\right),
\end{eqnarray}
and the neglected anti-rotating wave terms with the lowest frequency are terms oscillating with frequency of $2\omega_\text{tq}$, that is,
\begin{eqnarray} \label{ajc}
H_{\text{AJC}} &=& g \left(a^\dagger \sigma^+ e^{2i\omega_\text{tq} t}
+ a \sigma^- e^{-2i\omega_\text{tq} t}\right).
\end{eqnarray}
We therefore map  the effective model in Eq. (\ref{total4}) to the well-known Jaynes-Cummings model. This resonant interaction --- a bosonic quantum bus Hamiltonian --- is readily for quantum information transfer from a topological qubit to the cavity state~\cite{ZWY}. The first experiment may be the vacuum Rabi oscillation by preparing an initial state of $|\psi_1\rangle=(|0\rangle_\text{t} + |1\rangle_\text{t})/\sqrt{2}\otimes |0\rangle_\text{c}$, the quantum information transfer can be achieved by obtaining a final state of $|\psi_1\rangle_\text{f}=|0\rangle_\text{t}\otimes (|0\rangle_\text{c} - i|1\rangle_\text{c})/\sqrt{2}$ at $T_\text{g}=\pi/(2g)$, where the excitation of the topological qubit is transferred to the cavity mode. This dynamics can be directly probed in experiments. In this way, we can consolidate the advantage of both quantum systems in a single chip.

\bigskip

\noindent\textbf{Robustness of the  MF wavefunction.} The appearance of MFs at ends of the nanowire is ensured by the bulk topology. In this case, topological protected zero-energy edge states  can be realized at the two ends when the length of the nanowire $L \rightarrow \infty$. These localized edge states directly ensures self-hermitian, $\gamma = \gamma^\dagger$. The wave function of these edge states decays exponentially to zero in the bulk. For a finite system, the overlap of the two MF wave functions' tails lead to the MF energy splitting, which has been defined in Eq. (\ref{mf}). Here, as shown in Eq. (4), the decoherence of the topological qubit is originated from the fluctuation of hybridized energy splitting, and thus the stability of  MFs energy splitting against disorder means the robustness of the defined topological qubit against disorder. It is still not quite clear how robust this splitting is in a realistic nanowire because this energy splitting is in principle not topologically protected, and thus we can not directly infer its robustness from the topological protection. Nevertheless, robustness of this splitting is crucially important for the coupling between conventional and topological qubits. Therefore, we next investigate this problem using a tight-binding numerical simulation and a perturbation approach.

\begin{figure}[tb]
\begin{center}
\includegraphics[width=3.2in]{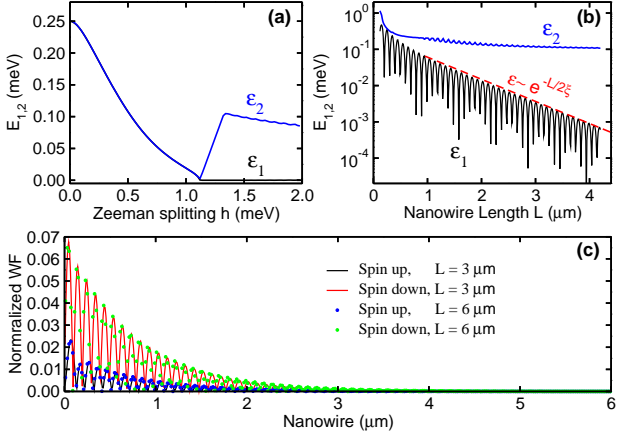}
\end{center}
\caption{\textbf{Lowest energy and wave functions of edge states in nanowires.} (a)  The topological transition in an infinity nanowire with open boundary, where $\varepsilon_1$ and $\varepsilon_2$ are the energy of two lowest particle levels. (b) Effect of nanowire length $L$ on $\varepsilon_1$ and $\varepsilon_2$ for fixed Zeeman field $h = 1.5$ meV. The solid line is the envelope of $\varepsilon_1$ fitted using $\exp(-L/2\xi)$ with $\xi = 230$ nm, which roughly agrees with the
Cooper pair coherent length $\xi_0 \sim 216$ nm.  (c) Normalized
wave function (WF) of the left end state, which is constructed from
$\psi_\text{L} \sim (\psi_{+1} + \Sigma^\dagger \psi_{+1})$ for different
nanowire length. The parity of this state is $+1$ because
$\Sigma^\dagger \psi_\text{L} = +\psi_{\text{R}}$. The right end state with parity
$-1$ has similar feature thus  is not shown. Other parameters from
InSb nanowires are: $m^* = 0.015m_0$, $\alpha = 20$ meV$\cdot$ nm,
$a = 10$ nm, $\Delta = 0.5$ meV and $\mu_{\text{eff}} = 1.0$ meV. The overlap
of these two wave functions with slightly different lengths is greater than
$0.99$.}
\label{fig-fig2}
\end{figure}

\emph{(1). Tight-binding simulation.} There are several sources of
fluctuation in nanowires, e.g., fluctuations of order parameters (the nanowire length is much larger the Cooper pair coherence) and chemical potential (small carrier density $n \sim 10^4$/cm), {\it etc}. To mimic these effects on the energy splitting, we consider the following tight-binding model
\begin{eqnarray} \label{tbh}
\mathcal{H} & =& -t \sum_{i,s} c_{is}^\dagger c_{i+1s} +  \lambda
\sum_{i} (c_{i,\uparrow}^\dagger c_{i+1,\downarrow}
-c_{i,\downarrow}^\dagger c_{i+1,\uparrow} + \text{h.c}) \nonumber\\
&&  + \sum_{i} \mu_{is} n_{is} + \Delta_i
c_{i\uparrow}^\dagger c_{i\downarrow}^\dagger + \text{h.c.},
\label{eq-TB}
\end{eqnarray}
where $t = \hbar^2/(2m^* a^2)$, $\lambda = \alpha/2a$ with $a$
the lattice spacing, $m^*$ is the effective mass of electron,
$\alpha$ is the spin-orbit coupling strength, chemical potential $\mu_{i\uparrow/\downarrow} =\mu_i \pm h$, and the Zeeman
splitting $h = g^*\mu_B B_\text{z}$ with $g^*$ being the Lande factor and
$B_z$ being the external magnetic field strength along the $z$-direction.  The topologically protected edge states appear when $h > h_\text{c} = \sqrt{\mu^2 + \Delta^2}$ with $\Delta$ being the proximity induced pairing strength, see Fig. \ref{fig-fig2}(a), which is protected by a finite energy around 0.1 meV for the parameters used therein. The energy splitting is an oscillation function of length $L$, see Fig. \ref{fig-fig2}(b). This is because the localized edge states have oscillating decay function, thus the overlap may exactly disappear at some "magic points". We plot the wave function of the left edge state in Fig. \ref{fig-fig2}(c) with different lengths, which shows that they are almost unchanged except their tails.

\begin{figure}[tb]
\begin{center}
\includegraphics[width=3.2in]{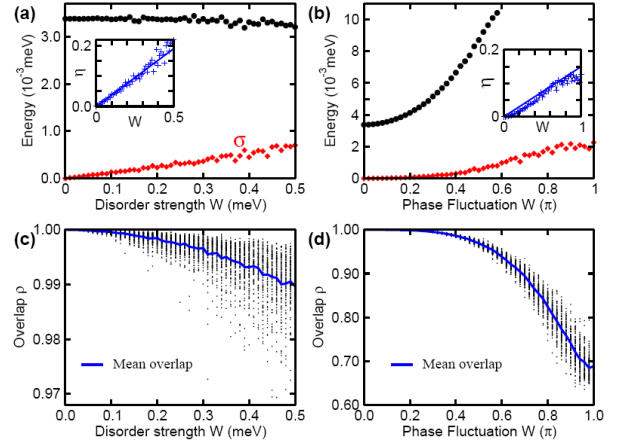}
\end{center}
\caption{\textbf{Effect of parameters fluctuations on the energy splitting.} Effect of chemical potential (left column) and phase (right column) fluctuations for a nanowire with $L = 3$ $\mu$m, $h = 1.5$ meV, and other parameters are the same as in  Fig. \ref{fig-fig2}. (a) and (b) show the fluctuation of chemical potential and phase on $\langle \varepsilon_1 \rangle$ and its variation
$\sigma = (\langle \varepsilon_1^2\rangle - \langle \varepsilon_1\rangle^2)^{1/2}$.
Insets show $\eta = \sigma /\langle \varepsilon_1\rangle$ vs $W$, and the
linear line is just for guide. The corresponding overlap of wave
function of MFs for  100 different fluctuations are presented in (c) and (d), with solid line the mean of overlaps. The overlap
is defined as $|\langle \psi_0| \psi_j\rangle|$, where $|\psi_0\rangle$ and $|\psi_j\rangle$ are the wavefunction of the edge states
without and with disorder, respectively; see more details in text.}
\label{fig-fig3}
\end{figure}

We now present our numerical  results by solving the tight-binding model over
different random configurations. Here we consider two typical fluctuations. In the first study, we assume on-site chemical potential fluctuation $\mu_i = \mu - 2t + \delta \mu_i$, and fix $\Delta_i = \Delta$. In the second study, we assume $\mu_i =
\mu$, and $\Delta_i = \Delta e^{i\delta \theta_i}$. These two factors are two major fluctuations in low dimensional systems with low carrier density.  In both cases, we assume $\delta \mu$ and $\delta \theta$ are independent uniform random numbers distributed in a large region $ [-W, W]$ (assuming $W \ge 0$). The results are presented in Fig. \ref{fig-fig3}, in which we mainly focus on the lowest two non-negative eigenvalues $\varepsilon_1$ and $\varepsilon_2$ of Hamiltonian in Eq. (\ref{tbh}). For the chemical potential fluctuation in Fig. \ref{fig-fig3}(a), the averaged Hamiltonian is exactly unchanged, thus we see the mean value of $\varepsilon_1$ is almost unchanged. We find that the variation of $\varepsilon_1$ almost increase linearly with respect to $W$. In Fig. \ref{fig-fig3}(c), we plot the overlap $\rho_j = |\langle \psi_0| \psi_j\rangle|$ as a function of $W$, where $|\psi_0\rangle$ is the wave function without disorder. Notice that the overlap of the left and right edge states is extremely small (at the order of $10^{-4}$ from our numerical simulation), thus $\rho \simeq 1$ means that the wave function of the edge state  is almost unaffected in a disordered environment. In the second column, we consider robustness of MFs with respect to the phase fluctuation. Two notable differences have been observed. First, as shown in Fig. \ref{fig-fig3}(b), the averaged Hamiltonian is changed because $\langle e^{i \delta\theta} \rangle \ne 0$, thus we find that $\langle \varepsilon_1\rangle$ depends strongly on $W$. Secondly,  as shown in Fig. \ref{fig-fig3}(d), we find that the overlap of the wave functions also depends strongly on the phase fluctuation magnitude. However, we have chosen extremely strong fluctuations in both cases, while  these fluctuations should be much smaller in realistic experiments. As a result, we may expect the practical performance to be much better  than the results presented in Fig. \ref{fig-fig3}. These simulations under the extreme condition  demonstrate clearly robustness of the MF wavefunctions. For this reason, we also expect that the topological qubits has a much weaker dephasing effect than conventional superconducting qubits do. It is worth to point out that we have also calculated the effect of nuclear spin polarization on the  energy splitting of MFs, where we have also obtained similar results. In our simulation, we assume a random magnetic field generated by nuclear spins, ${\bf B}_i$, which introduce a Zeeman splitting smaller than 0.1 meV.

\emph{(2) Perturbational analysis.} We now develop a model to understand these numerical observations. We wish to show that robustness of MFs splitting not only deeply rooted in the bulk topology. To this end, we assume $H = \mathcal{H} + V$, where $\mathcal{H}$ is the unperturbed model defined in Eq. (\ref{eq-TB}) and $V$ is the disordered potential, which contains all possible random fluctuations. This model has the basic particle-hole symmetry, that is, $\Sigma = \sigma_\text{x} K$, where $K$ is the complex conjugate. Now we assume $\mathcal{H} \psi_n =\varepsilon_n\psi_n$, then $\mathcal{H} \Sigma^\dagger \psi_n =-\varepsilon_n \Sigma^\dagger \psi_n$. Hereafter, for convenience, we assume $n > 0$ and $n < 0$ for the eigenfunctions with position and negative eigenvalues, respectively, and thus $\varepsilon_{+n} = -\varepsilon_{-n}$, $\psi_{-n} = \Sigma^\dagger \psi_{+n}$. The system is protected by a fundamental gap, see Fig. \ref{fig-fig2}(a), which is in the order of  magnitude $\Delta$. We attempt to understand the topological protection using the second-order perturbation theory. To this end, we assume the two localized wave functions as $\psi_{\text{L}}$ and $\psi_{\text{R}}$, where the subscript L (left) and R (right) represent the position of the end states; see a typical example in Fig. \ref{fig-fig2}(c). These two edge states have the following basic features:
$\Sigma^{\dagger} \psi_{\text{L}}  = \psi_{\text{L}}$ and
$\Sigma^{\dagger} \psi_{\text{R}}  =  - \psi_{\text{R}}$.
The eigenfunction of $\psi_{\pm 1}$ can be constructed using the above two edge states as
$\psi_{\pm 1} = \mathcal{A} (\psi_{\text{L}} \pm \psi_{\text{R}})$,
where $\mathcal{A} \simeq 1/\sqrt{2}$ is the renormalization constant, and $\psi_{\pm 1}$ is the eigenfunction of $\mathcal{H}$ with eigenvalue $\varepsilon_{\pm 1}$.

The random potential $V$ can affect the low-lying excitation. To this end, we assume
$\epsilon_i' =\epsilon_i + \delta \epsilon_i^{(1)} + \delta \epsilon_i^{(2)}  + \cdots$,
where the first order correction is
\begin{equation}\label{eq-LR}
\delta \epsilon_{\pm 1}^{(1)} = \langle \psi_{\pm 1} | V | \psi_{\pm 1} \rangle = 2 \Re \langle \psi_\text{L} |V| \psi_\text{R} \rangle.
\end{equation}
The above conclusion can be obtained using the following identity:
$\langle \psi_\text{k} |V| \psi_\text{k} \rangle = \langle \psi_\text{k} |\Sigma^\dagger V\Sigma| \psi_\text{k} \rangle = -\langle \psi_\text{k} |V| \psi_\text{k} \rangle$ with $k = \text{L, R}$,
and thus $\langle \psi_\text{k} |V| \psi_\text{k} \rangle \equiv 0$ for any weak random potential. As the wavefunction of the left and right edge states --- ensured by topology --- is an exponential decay function, see Fig. \ref{fig-fig2}(c), their overlap, due to the random potential $V$, should be exponentially decay to zero with increasing the length, that is, $\delta \epsilon_{\pm 1}^{(1)} \sim \exp({-L/\xi})$. Generally, we find  $\delta \epsilon_{\pm 1}^{(1)}/\epsilon_1 \ll 1$. This small ration arises from the oscillation of the edge state wavefunction, in which most of the important contributions are exactly cancelled. In contrast, for conventional qubits, the first-order fluctuations play normally the leading role in the energy fluctuation (thus the decoherence) of qubits.

We next calculate the second-order correction energy, which can be written as (see the Method section for details),
\begin{eqnarray}\label{2nd}
\delta \epsilon_{+1}^{(2)} = \sum_{n>1} {l_{n}r_n + l_{-n}r_{-n} \over
\epsilon_n - \epsilon_{+1}} + {2 |\Im \langle \psi_\text{L} |V| \psi_\text{R} \rangle |^2 \over
\epsilon_{+1}}.
\end{eqnarray}
where $l_n = \langle \psi_\text{L}|V| \psi_n\rangle$, $r_n =\langle \psi_n|V| \psi_\text{R}\rangle$, and no correlation between them can be derived for a general random potential $V$. However, the most important contributions of $l_n$ and $r_n$ to $\delta \epsilon_{+1}^{(2)}$ are almost cancelled due to the particle-hole symmetry. Notice that the first term in the above equation is suppressed by the large energy gap, i.e., $\epsilon_n - \epsilon_{+1} \sim \Delta$, and thus the cross correction between the left and right edge states is negligible when $V$ is not very strong in a finite length system. Meanwhile, the second term is also exponentially small when $L \gg \xi$ due to the exponential small overlap between the wavefunctions of the two edge states. Finally,  this result is also in consistent with the bulk-edge correspondence in quantum phase transition since when $L \rightarrow 0$, we see $\delta \epsilon_{+1}^{(2)} \rightarrow 0$, that is, the energy of the edge states are unaffected by $V$.

From these results, we can conclude that the topological qubit, even with a finite coupling, is still robust against local perturbations --- a basic reason relies on the topology of the bulk. Following these observations, we expect the topological qubit has much smaller dephasing rate, which can be regarded as one major advantage of it. In addition, these results are quite general, and for other topological qubits with some other symmetries, we also expect a similar conclusion.

\bigskip
\noindent\textbf{Realization of the quantum information transfer.} Topological qubits embedded in an environment inevitably have the finite lifetime ---this process can be modelled by two parameters: the relaxation rate $\Gamma_1$ and the dephasing rate $\Gamma_2$. Here we investigate these two main decohenrence sources for practical experimental realizations of the current scheme. First, dissipative and incoherent quasiparticles tunneling across the Josephson
junction and between nanowire and superconductors will break the parity of the qubit system and lead to decoherence.  At a temperature of 20 mK \cite{25mk}, the density of unpaired quasiparticles is 0.04 $\mu$m$^{-3}$, which leads the parity protection time to be in the order of magnitude $\sim$1 ms \cite{qp}. This is sufficiently large comparing with the time for the quantum information transfer process.
Secondly,  we consider the influence of the superconducting phase fluctuation effect on the MF coupling, which  comes from the thermal
fluctuations of the bias voltage. For the free $\sigma_x$  term, it is a fast oscillation one, and thus low frequency modulations
of the term can be negligible provided that the frequencies are much smaller comparing with $\omega$.  As for the other terms,
random superconducting phase fluctuations do affect the form of MFs, thus leads to decoherence of the topological qubit. The
root mean square error of the superconducting phase is $\delta\phi_{\text{rms}}\simeq A/e \leq 10^{-3}$ \cite{Hassler11} with $A
\in [10^{-4}, 10^{-3}]e$ \cite{A} being the amplitude of the $1/f$ charge noise, the error caused by which is negligible small and
far below the threshold \cite{imp1,imp2} for error correction.

We now discuss realistic parameters. In circuit QED \cite{rms}, the resonator has a  wave length  of $\lambda=25$ mm and a gap of
$d=5$ $\mu$m between the center conductor and its ground planes, being large enough for a transmon qubit with the loop size of
$4\times 4$ $\mu$m$^2$.  The transmission probability of the junction is very small, which depends on $l_1$ and the magnetic
field \cite{acmh}, and thus can be tuneable by tuning  $l_1$ with external electrostatic gates \cite{gate}. We modulate
$E_\text{M}=2\pi \times 0.5$ GHz, thus $g\approx 2\pi\times 6$ MHz for $\lambda_\text{c}/\omega_\text{c}\simeq 0.05$ \cite{g}.  Quality factors above one million have been reported for superconducting resonators  with frequencies ranged
from 4 to 8 GHz \cite{decay}, i.e., the cavity decay rate $\kappa$ in the order of KHz. Here we choose $\kappa=2\pi\times 6$ KHz, which corresponds to
$\kappa=g/1000$. Meanwhile, we choose $E=E_2=2\pi\times 0.2$ GHz, which leads to $l_2\approx2.5$ $\mu$m  for an InSb wire with $\xi= 216$ nm.
For $l_0=4.5$ $\mu$m, $E_1$ will be less than $0.01E$, and thus
can be safely neglected. Therefore, the total length of the wire
will be less than 10 $\mu$m,  which can be deposit on a transmon
qubit. In addition, as $\omega_\text{tq}= E$, the resonate condition $\omega_\text{tq}+\omega=\omega_\text{c}$ is readily fulfilled with  $\omega=10 E_\text{M}=2\pi \times 5$ GHz and  $\omega_\text{c}=2\pi \times 5.2$ GHz. Obviously, these parameters naturally realize strong coupling between topological qubits and cavity since $g \gg \{\kappa, \Gamma_1, \Gamma_2\}$.

\begin{figure}[tb]
\begin{center}
\includegraphics[width=3.4in]{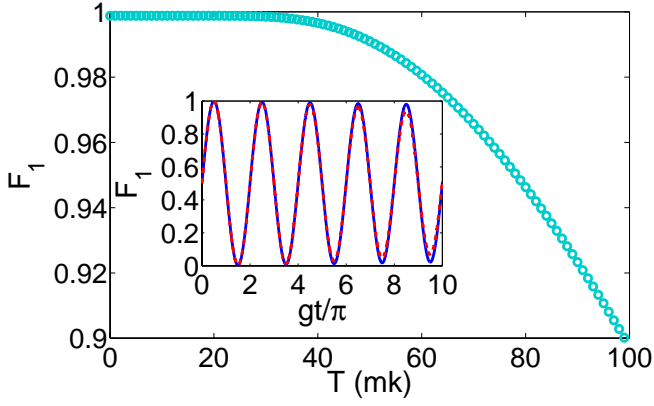}
\end{center}
\caption{\textbf{Fidelity of the quantum information transfer between the topological qubit and the cavity.} The maximum fidelity  as a function of the
working temperature $T$ of the cavity  with $\kappa=\Gamma_1=\Gamma_2=g/1000$.  Insert: The fidelity dynamics as a function of $gt/\pi$ at zero temperature. The blue and red dash lines are simulated with and without rotating wave approximation,
respectively.}
\label{f1}
\end{figure}

We estimate the errors for the quantum information transfer process under realistic conditions. First, we consider the decay of the cavity mode ($\kappa$) with a thermal cavity photon number $n$, the relaxation ($\Gamma_1$) and  dephasing ($\Gamma_2$)  of  the topological qubit.  Under these decoherence effects, the dynamics of the system can be well described by the  master equation
\begin{eqnarray}\label{master}
\dot\rho_1 &=& -i[H_{\text{JC}}, \rho_1]
+\frac \kappa  2 \left[(n_\text{c}+1) \mathcal{L}(a)+ n_\text{c} \mathcal{L}(a^\dag)\right] \notag\\
&& + \frac 1 2 \left[\Gamma_1 \mathcal{L}(\sigma^-) + \Gamma_2 \mathcal{L}(\sigma^\text{z}\right],
\end{eqnarray}
where $\rho_1$  is the density matrix of the combined system of
the topological qubit and the  cavity, $n_\text{c}$ is the number of photon in the cavity, and $\mathcal{L}(A)=2A\rho A^\dagger-A^\dagger A \rho -\rho A^\dagger A$ is the Lindblad operator.  We simulate the quantum information transfer process using the conditional fidelity defined by $F_1=_\text{f}$$\langle\psi_1|\rho_\text{a}|\psi_1\rangle_\text{f}$, with $\rho_\text{a}$ being the reduced  density matrix of the topological
qubit from $\rho_1$. Assuming the cavity is initially prepared in
the vacuum state $|0\rangle_\text{c}$, i.e., $n_\text{c} =0$, we obtain a high  fidelity of $F_1 \simeq 99.9\%$  for the quantum information transfer process at $gt/\pi=1/2$ with $\kappa=\Gamma_1=\Gamma_2=g/1000$.

We next turn to consider the influence when  the cavity is  initially
in a  thermal state. Typically, the cavity is cooled down near its
quantum mechanical ground state and the thermal occupancy related
to the working temperature T of the cavity as
$n_\text{c}=1/[\exp(\hbar\omega_\text{c}/K_\text{B} T)-1]$. To simplify our treatment, we assume the initial state  of the thermal cavity to be
$\rho_\text{ic}=(1-n_\text{c})|0\rangle_c\langle 0|+n_\text{c}|1\rangle_c\langle 1|$. With the same parameters as above, as shown in Fig. \ref{f1}, we
plot the maximum of $F_1$, with rotating wave approximation, as a
function of T.  We find that the infidelity  is less than 0.1\%
when $\text{T}\leq 35$ mK. For superconducting devices cooled to 20 mK inside a dilution refrigerator, the temperature effect in our scheme is negligible.

Finally, we consider the  influence of neglecting the counter-rotating terms in deriving the Hamiltonian in Eq. (\ref{jc}). Here the neglected terms with frequencies in the order of $\omega_\text{tq}$ are those in the Hamiltonian of Eq. (\ref{ajc}). This is well justified numerically for $\omega_\text{tq}=E\approx 33 g$, as shown by the insert of  Fig. \ref{f1},  where the blue and red dashed lines are simulated by the Hamiltonian of $H_\text{JC}$  in Eq. (\ref{jc}) with the absence and presence of $H_\text{AJC}$  in Eq. (\ref{ajc}), respectively. The two results are in very good agreement,   and the infidelity induced by this approximation  is less than 0.1 \% within the three periods of Rabi oscillation.

\bigskip
\noindent\textbf{Application to entangled states generation.} When incorporating more than one qubit, we next show  that our quantum bus model can be naturally used to generate entangled states of topological qubits.  We consider the multi-qubit case as shown in Fig. \ref{qubit}(b) and modulate
$\nu=\omega_\text{c}-E_j-\omega_j>0$ for all the $N$ qubits, which leads
the total interaction Hamiltonian to be
\begin{eqnarray}\label{st}
H_\text{MF}^{\text{N}}=g\sum_{j=1}^N\left(a e^{-i\nu t}\sigma^+_j+\text{H.c}. \right),
\end{eqnarray}
where we have assumed $g_j=g$. Meanwhile, driving in  the form of
$h_\text{D}=\epsilon a^\dagger e^{-i\omega_\text{d} t}+ \epsilon^* a
e^{i\omega_\text{d} t}$ on the resonator can be obtained \cite{rms} by
capacitively coupling it to a microwave source with frequency
$\omega_\text{d}$, with $\epsilon$ being a time independent amplitude. For
large amplitude driving and under a  time-dependent displacement
transformation of $D(\alpha) = \exp\left(\alpha_\text{d} a^\dagger -
\alpha^*_\text{d} a \right)$ with $i\dot\alpha_\text{d} = \omega_\text{c}\alpha_\text{d} +\epsilon \exp({-i\omega_\text{d} t})$, the direct drive on the resonator can be eliminated. Under resonant driving
($\omega_d=\omega_\text{tq}$), and change to a frame rotating at the frequency of $\omega_\text{d}$, the driven induced  collective Rabi oscillating Hamiltonian of the topological qubits   reads $H_\text{D} = {\Omega\over 2} \sum_{j=1}^N
\sigma^\text{x}_j$, where $\Omega=2g\epsilon/\nu$. In the interaction picture with respect to $H_\text{D}$, the interaction
Hamiltonian reads \cite{s}
\begin{eqnarray}
H_{\text{MF}}^\text{N} &=& {g \over 2} \sum_{j=1}^N  a e^{-i\nu t} \left(\sigma^\text{x}_j +e^{i\Omega t}|+\rangle_j\langle-|
-e^{-i\Omega t}|-\rangle_j\langle+|\right) \notag\\ &&+ \text{H.c},
\end{eqnarray}
where $|\pm\rangle_j=(|0\rangle_j\pm|1\rangle_j)/\sqrt{2}$. In the
case of $\Omega\gg\{\nu, g\}$, we can omit the fast oscillation
terms (of frequencies $\Omega\pm\delta$), then the Hamiltonian
becomes
\begin{eqnarray} \label{h2}
H_{\text{MF}}^\text{N} =g \left(a e^{-i\nu t}+a^\dag e^{i\nu t}\right)
J_\text{x},
\end{eqnarray}
where $J_\mu=\Sigma_{j=1}^N \sigma_j^\mu/2$ with $\mu=\text{x, y, z}$. In
this case, the time evolution operator can be written as \cite{Zhuprl20031,Zhuprl20032,ms1,ms2}
\begin{eqnarray}
\label{un}
U(t)=\exp\left[{i A(t) J_\text{x}^2}\right] \exp\left[{iB(t)a J_\text{x}}\right]
\exp\left[{iB^*(t)a^\dag J_\text{x}}\right],\notag\\
\end{eqnarray}
where
$B(t)=ig \left(1-e^{-i\nu t}\right)/\nu$, and
$A(t)={g^2 \over \nu} \left[t + i\left(e^{i\nu t}-1\right)/ \nu\right]$.
It is obvious that $B(t)$ is a periodical function and equals zero when $t =
T_k=2k\pi/\nu$  with $k$ being a  positive integer. Therefore, at
these special points,  Eq. (\ref{un}) reduces to
\begin{eqnarray}\label{ut}
\label{unreduce} U(T_k)=\exp\left[i A(T_k) J_\text{x}^2\right],
\end{eqnarray}
where $A(T_k)=2k\pi g^2/\nu^2$. For $N$ qubits in an initial state of  $|\psi_2\rangle_\text{i}=|00\cdots 0\rangle$, choosing $A(T_k) =\pi/2$, the final state $|\psi_2\rangle_\text{f}=\exp{\left(i\frac{\pi}{2}J_x^2\right)}|\psi_2\rangle_\text{i}$ is  a GHZ state given by \cite{ms1,ms2}
\begin{equation}\label{GHZ}
|\psi_2\rangle_\text{f}=\frac{1}{\sqrt{2}}[|00\cdots 0\rangle +e^{-i\pi(1+ N)/2}|11\cdots
1\rangle],
\end{equation}
when $N$ is even. For odd $N$, one  can get GHZ state by applying
$U_\text{D}=\exp(i{\pi \over 2} J_\text{x})$ in
addition to Eq. (\ref{ut}). The operator $U_\text{D}$ can be implemented
by $H_\text{D}$ with $\Omega T_k=3\pi$.

This generation has the following  distinct merits. First, the
generation is fast.  To be specifically, $A(T_k) =\pi/2$ can be obtained when $\nu=2g\sqrt{k}$. Then, for $k=1$, one obtains $\nu=2g$ and the entanglement generation time  $T=\pi/g$, which is comparable with that of using the resonant Jaynes-Cummings interaction. This is due to a fact that the interaction used in this generation is not of the dispersive nature, and thus removes the needs of large
detuning ($\nu \gg g$). Secondly, the generation is readily for
scale up. As the operator in Eq. (\ref{ut}) is obtained to be
independent on the number of the involved qubits,  the time
needed for the gate operation does not depend on the number of
qubits. Therefore, this generation can be scalable provided that the
qubits can be incorporated in the cavity for every wave length
section of the cavity, and there can be four qubits located at the
antinodes, as shown in Fig. \ref{qubit}. Finally, in the time
evolution operator of Eq. (\ref{un}), as $B(t)$ is a periodical function, the
cavity state dependent terms, i.e., the second and third terms,
are removed, leading to a cavity field state insensitive
operator of Eq. (\ref{ut}). Since the cavity will return to its
original state, one can avoid cooling of the cavity to its ground
state before the application of the operator in Eq. (\ref{ut}), which looses the
limitation of the thermal effect in engineering quantum states.

\begin{figure}[tb]
\begin{center}
\includegraphics[width=3in]{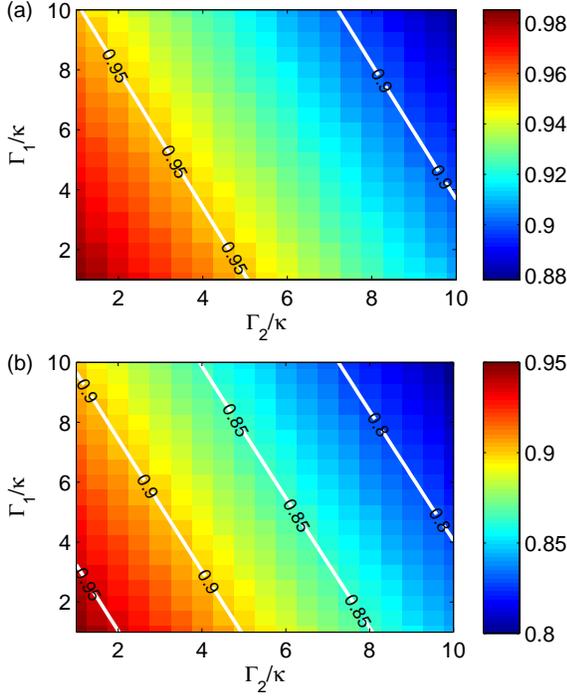}
\end{center}
\caption{\textbf{Maximum of the entanglement generation fidelity $F_2$.} For (a) $N=4$ and (b) $N=8$,  $F_2$ is  plotted at  $T=0$ with $\kappa=g/1000$  and $\{\Gamma_1, \Gamma_2\} \in [1, 10]\kappa$.}
\label{f2}
\end{figure}

However,   the time evolution does involve the excitation of the cavity during the generation, so that we need to include its influence as
well as others. Then, we estimate the fidelity for the generation
process  by the Lindblad master equation. For the $N=2$ case,  we
can  obtain a high fidelity of $F_2\simeq 99.3\%$ at $t=\pi/g$ for
the generation with $\kappa=\Gamma_1=\Gamma_2=g/1000$ at $T=0$.
For $N>2$ cases, the maximum of $F_2$ will decrease gradually  due
to the decoherence of the increased number of qubits.
Nevertheless, we can still obtain fidelities of $98.5\%$ and
$96.8\%$ for the entanglement generation with $N=4$ and $N=8$,
respectively. As it is well known, the  fidelity of the generation
drops with the increase of the decohenrence rates.  For
the cases of $N=4$ and  $N=8$, as shown in Fig. \ref{f2}, we also
plot the maximum of $F_2$ with decohenrence rates  in the range of
$\Gamma_{1,2} \in [1, 10]\kappa$. It should be emphasized that the dephasing term $\Gamma_2$ has a leading effect in $F_2$ for the multipartitie entangled state. In the previous sections, we demonstrate that the topological qubits is much more stable than the conventional qubits in environment, and thus we expect $\Gamma_2$ to be much smaller than that in conventional qubits, namely, $F_2$ for topological qubits can be much higher than that for conventional qubits.

In summary,  we have proposed a microwave photonic quantum bus for a direct coupling between flying and topological qubits, in which the energy mismatch is compensated by the external driving field. Strong coupling between these two qubits can be realized. It has also been shown that from the realistic tight-binding simulation and perturbation theory that the energy splitting of the MF wavefunctions in a finite length nanowire is still robust against local perturbations, which is ensured by the topology. Thus our scheme is rather promising for implementing a robust interface between the flying and topological qubits. Finally, we have demonstrated that this quantum bus can be used to generate multipartitie entangled states with the topological qubits.

\bigskip
\noindent\textbf{Method\\}
\textbf{Derivation of Eq. (\ref{total6}).}
We begin with the Hamiltonian in Eq. (\ref{total4}) in the main text. Using the  series  identities of
\begin{eqnarray}
\cos ( \theta\cos\omega t)&=&J_0(\theta)-2\sum_{n=1}^\infty J_{2n}( \theta) \cos(2n\omega t)\notag\\
&\equiv& J_0-2J_\text{E}(\theta),
\end{eqnarray}
and
\begin{eqnarray}
\sin ( \theta\cos\omega t)&=& 2\sum_{n=1}^\infty (-1)^{n+1}
J_{2n-1}( \theta) \cos[(2n-1)\omega t]\notag\\
&\equiv&2J_\text{O}(\theta)
\end{eqnarray}
with  $J_n(\theta)$ being the  $n$th Bessel function of the first kind,  the Hamiltonian in Eq. (\ref{total4})  reads
\begin{eqnarray}\label{h1}
H_\text{MF}&=&{E \over 2} \sigma^\text{z} -{E_\text{M} \over 2}
\cos\varphi_0 J_0\left(\theta\right)\sigma^\text{x} +S(t) \notag\\
&& -g_0\sigma^\text{x} K(t) \left(a e^{-i\omega_\text{c} t}
+a^\dag e^{i\omega_\text{c} t}\right),
\end{eqnarray}
where we have defined the time-dependent driven as
\begin{eqnarray}
S(t)&=&E_\text{M}[\sin\varphi_0 J_\text{O}(\theta)
+ \cos\varphi_0 J_\text{E}(\theta)]\sigma^\text{x},\notag\\
K(t)&=&\left[\sin\varphi_0 J_0\left(\theta\right)
+ 2\cos\varphi_0 J_\text{O}(\theta) -2 \sin\varphi_0 J_\text{E}(\theta)\right]. \notag
\end{eqnarray}

To obtain a time-independent effective Hamiltonian for Eq. (\ref{h1}), we first need to deal with the time-dependent driven terms of $S(t)$. This time-dependency can be safely neglected when $J_n(\theta)E_\text{M}/(n\omega)\ll 1$. To see this, we perform $n$ transformations with frequencies $n\omega$, which are defined by $U(t)=U_nU_{n-1}...U_2U_1$ with
\begin{eqnarray}
U_n&=&\exp\left[i \beta_m \sin  (m\omega t) \sigma^\text{x}\right],
\end{eqnarray}
where $\beta_m =(-1)^{(m-1)/2}\sin \varphi_0 E_\text{M} J_m(\theta) / (m\omega)$ and $\beta_m =\cos \varphi_0  E_\text{M} J_m(\theta) / (m\omega)$ for odd $m=2n-1$  and even $m=2n$, respectively. The transformed Hamiltonian is
\begin{eqnarray}\label{trans}
H_{\text{MF}}&=&U(t)H_\text{MF}U^\dagger(t)-i U(t) \frac{\partial U^\dagger(t)} {\partial t},
\end{eqnarray}
where the second term equals to $S(t)$, and thus cancels the time-dependency of $S(t)$ in Hamiltonian (\ref{h1}). However, the $\sigma_\text{z}$ term does not commute with the transformation.  After $n$ transformations, its transformed form is $T_n=U_nT_{n-1}U^\dagger_n$, where
\begin{eqnarray}
T_1&=&U_1\sigma_\text{z}U^\dagger_1
=\cos[2 \beta_1 \sin  (\omega t)]\sigma_\text{z}
+\sin[2 \beta_1 \sin  (\omega t)]\sigma_\text{y}\notag\\
&=& \left[J_0(2\beta_1)+2\sum_{n=1}^\infty J_{2n}( 2\beta_1)
\cos(2n\omega t)\right] \sigma_\text{z} \notag\\
&+&2\sigma_\text{y} \sum_{n=1}^\infty  J_{2n-1}(2 \beta_1 )
\cos[(2n-1)\omega t].
\end{eqnarray}
Choose $\theta=0.4$ leads to $J_1(\theta)\approx \theta/2=0.2$, and thus
$$\beta_1= \sin \varphi_0 {E_\text{M} J_1(\theta) \over \omega}
\leq  {E_\text{M} J_1(\theta) \over \omega}
\approx {E_\text{M} \over 5\omega} = 1/50,$$
for $\omega=10 E_\text{M}$. Therefore, $J_0(2\beta_1)\geq 0.9996$, $J_1(2\beta_1) < 0.02$ and $J_n(2\beta_1)<0.0002$ with $n\geq2$, and thus
\begin{eqnarray}\label{trans1}
T_1&\approx& J_0(2\alpha_1) \sigma_\text{z} \approx \sigma_\text{z}.
\end{eqnarray}
Similarly, as $J_2(\theta)\approx 1/50$, $$\beta_2= \cos \varphi_0 {E_\text{M} J_2(\theta) \over 2\omega}\leq  {E_\text{M} \over 100\omega}=0.001,$$
and thus $J_0(2\beta_2)=1$ and $J_n(2\beta_2)<0.001$ with $n\geq1$. Therefore,
\begin{eqnarray}\label{trans2}
T_2=U_2 T_1 U_2^\dagger \approx   U_2  \sigma_\text{z} U_2^\dagger
\approx J_0(2\beta_2) \sigma_\text{z}= \sigma_\text{z}.
\end{eqnarray}
As $\beta_n\ll \beta_2$ for $n\geq 3$, which leads to $J_1(2\beta_n)\ll J_1(2\beta_2) <0.001$, and thus
\begin{eqnarray}\label{transn}
T_n\approx T_1\approx \sigma_\text{z},
\end{eqnarray}
which means  that $S(t)$ does not contribute to the effective Hamiltonian, thus can be safely neglected.

Therefore, neglecting $S(t)$, the Hamiltonian in Eq. (\ref{h1}) reduces to
\begin{eqnarray}\label{h2}
H_\text{MF} &=&{E \over 2} \sigma^\text{z}
-\cos\varphi_0 J_0\left(\theta\right) {E_\text{M} \over 2}\sigma^\text{x}\notag\\
&&-g_0 \sigma^\text{x} K(t)  \left(a e^{-i\omega_\text{c} t}+a^\dag e^{i\omega_\text{c} t}\right).
\end{eqnarray}
It is obvious that the energy splitting of the  topological qubit is $\omega_\text{tq}=\sqrt{E^2+(\cos\varphi_0 J_0\left(\theta\right) E_\text{M})^2}$. Usually, $\omega_\text{tq}$ is much smaller than $\omega_\text{c}$, one should use the external driven force, denotes by $K(t)$, to match this energy difference. To be more specifically, we rewrite  Hamiltonian (\ref{h2}) as
\begin{eqnarray}\label{h3}
H_\text{MF}&=&{E \over 2} \sigma^\text{z}
-\cos\varphi_0 J_0\left(\theta\right) {E_\text{M} \over 2}\sigma^\text{x}\notag\\
&-&g_0  J_0(\theta) \sin\varphi_0  \left(a e^{-i\omega_\text{c} t}+a^\dag e^{i\omega_\text{c} t}\right) \sigma^\text{x}\notag\\
&+&g_0 \sin\varphi_0 \sigma^\text{x} \sum_{n=1}^\infty J_{2n}(\theta)
\left[a^\dag e^{i(\omega_\text{c}-2n\omega) t} +h.c. \right] \notag\\
&-&g_0 \cos\varphi_0 \sigma^\text{x} \sum_{n=1}^\infty (-1)^{n+1} J_{2n-1}(\theta)\notag\\
&& \times \left[a^\dag e^{i(\omega_\text{c}-(2n-1)\omega) t} +h.c. \right],
\end{eqnarray}
where we have neglected the terms oscillating with frequencies $\omega_\text{c}+n\omega$. Therefore, resonate coupling can be induced when $\omega_\text{c}-n\omega=\omega_\text{tq}$ with the coupling strength $\sim g_0 J_n(\theta)$. As the coupling strength is proportional to $J_n(\theta)$, it will be relatively small when $n\geq 2$. Therefore, we consider $n=1$ case, i.e.,  $\omega_\text{tq}+\omega=\omega_\text{c}$. In this case, we can see from Hamiltonian (\ref{h3}) that one can  keep only $n=1$ term in $K(t)$. In addition, to obtain the maximum coupling strength, we choose $\varphi_0=\pi$, which leads Eq. (\ref{h3}) to Eq. (\ref{total6}) in the main text.

At this stage, we recheck the condition of $J_n(\theta)E_\text{M}/(n\omega)\ll 1$ in order to neglect $S(t)$. The choice of  $\varphi_0=\pi$ leads to $\beta_{2n-1}=0$, and thus we only need to ensure that $J_1(2|\beta_{2n}|)\ll 1$. Then, it is sufficient to require that $J_1(2|\beta_{2}|)\approx|\beta_{2}|\ll 1$. As $|J_2(\theta)|< 1/2$ for arbitrary $\theta$,
$$|\beta_{2}|=\left|{E_\text{M} J_2(\theta) \over 2\omega}\right| =\left|{J_2(\theta) \over 20}\right|< {1\over 40} \ll 1.$$
Therefore, in the case of $\varphi_0=\pi$ and $\omega/E_\text{M}=10$, there is no specific limitation with respect to $\theta$.

\bigskip
\noindent\textbf{Calculation of $\delta \epsilon_{\pm 1}^{(2)}$.}
First, we can expect that the second-order correction energy to $\psi_{\pm 1}$ is exactly equals to zero when $L \rightarrow \infty$. It is easy to understand from the following identity ($k = \text{L, R}$),
\begin{equation}
{\langle \psi_k |V| \psi_n \rangle \langle \psi_n | V | \psi_k \rangle \over \varepsilon_n} + {\langle \psi_k |V| \psi_{-n} \rangle \langle \psi_{-n} | V | \psi_k \rangle \over \varepsilon_{-n}} =0,
\label{eq-B1}
\end{equation}
where we have assumed that the left and right edge states $\psi_\text{L,R}$ have
well-defined chirality. Note that $\psi_\text{L,R}$ are not necessary to be
the eigenvectors of the Hamiltonian. Physically, it means that the second-order correction energy from the particle and hole sectors exactly cancels with each other, and thus $\delta \epsilon_{\pm}^{(2)} = 0$ when $L \rightarrow \infty$.

In the following, we wish to show that in the finite length case, contributions from the particle and hole sectors will also almost be canceled, and thus the net second-order correction energy is also very small. To this end, we need to calculate
\begin{eqnarray}
\delta \epsilon_{+1}^{(2)} =
&& \sum_{n > 1} \left({\langle \psi_\text{L} + \psi_\text{R} |V| \psi_n \rangle \langle \psi_n|V| \psi_\text{L} + \psi_\text{R} \rangle \over \varepsilon_n - \epsilon_{+1} } \right. \nonumber \\
&& \left.+ {\langle \psi_\text{L} + \psi_\text{R} |V| \Sigma^\dagger \psi_n \rangle \langle \psi_n| \Sigma V| \psi_\text{L} + \psi_\text{R} \rangle \over -\varepsilon_n - \epsilon_{+1} }\right) \nonumber \\
&& + {|\langle \psi_\text{L} + \psi_\text{R} |V| \psi_\text{L} - \psi_\text{R}\rangle |^2 \over \varepsilon_{+1} - \varepsilon_{-1}}.
\end{eqnarray}
The correlation energy to $\psi_{-1}$ can be calculated using a similar manner, and we can prove exactly that $\delta \epsilon_{+1}^{(2)} = - \delta \epsilon_{-1}^{(2)}$, which ensures that the perturbation method also respects the particle-hole symmetry.

Using the identity Eq. (\ref{eq-B1}), we obtain the correction energy  as in Eq. (\ref{2nd}), the matrix elements have the following general properties:
$\langle \psi_\text{L} | V | \psi_n\rangle = -(\langle \psi_\text{L} |V| \Sigma \psi_n\rangle)^*$ and
$ \langle \psi_\text{R} | V | \psi_n\rangle = +(\langle \psi_\text{R} |V| \Sigma \psi_n\rangle)^*$.
Notice that $\psi_n$ may contain an arbitrary phase, thus both $\langle \psi_\text{L} | V | \psi_n\rangle$ and $\langle \psi_\text{R} | V | \psi_n\rangle$
are generally complex numbers. In other words, the first term in Eq. (\ref{2nd}) is in general non zero. In fact, the second-order correction is exactly equal
to zero only when the left and right wavefunctions are well separated. In this case, $\psi_\text{L}$ and $\psi_\text{R}$ are also the eigenvectors of the
Hamiltonian, and thus $\langle \psi_\text{L} | V| \psi_\text{R}\rangle = 0$ and $\langle \psi_\text{L} | V| \psi_n\rangle = 0$.

The numerical results show that the contribution of the first term in Eq. (\ref{2nd}) is much smaller than  the second-term in a finite length system. This can be understood as follows. First, the system protected by a large energy gap, so the second-order contribution is greatly suppressed. Secondly, the edge states are fast oscillating function in real space, while the extended states $\psi_n$ are well-extended in the real space. Thus the overlap between the localized state and extended state mediated by the random potential is very small.  Therefore, the major contribution to the second-order correction energy comes from the second term in Eq. (\ref{2nd}). Note that $\langle \psi_\text{L}|V|\psi_\text{R}\rangle \sim \exp[-L/(2\xi)]$ and $\epsilon_{+1} \sim  \exp[-L/(2\xi)]$, thus it is reasonable to expect that the second term is also very small. Obviously, $\lim_{L\rightarrow \infty} \epsilon_{+1} (L) = 0$, which is in consistent with the well-known bulk-edge correspondence in topological phase transitions.

\bigskip

\noindent \textbf{Acknowledgements}\\
This work was supported by the NFRPC (No. 2013CB921804 and No. 2011CB922104),  the NSFC (No. 11125417, No. 11104096, and No. 11374117), the PCSIRT  (No. IRT1243), the GRF (No. HKU7045/13P and No. HKU173051/14P), and the CRF (No. HKU-8/11G)  of the RGC of Hong Kong. M.G. is supported by Hong Kong RGC/GRF Projects (No. 401011 and No. 2130352), University Research Grant (No. 4053072)  and The Chinese University of Hong Kong (CUHK) Focused Investments Scheme.

\end{document}